\documentclass[usenatbib]{mn2e}

\usepackage{epsf}

\newcommand\be{{\bmath e}}
\newcommand\bu{{\bmath u}}
\newcommand\bcdot{{\bmath\cdot}}
\newcommand\bnabla{{\bmath\nabla}}
\newcommand\bfT{\mathbf{T}}
\newcommand\bfone{\mathbf{1}}

\newcommand\half{{\textstyle\frac{1}{2}}}

\newcommand\rmb{\mathrm{b}}
\newcommand\rmd{\mathrm{d}}
\newcommand\rme{\mathrm{e}}
\newcommand\rmi{\mathrm{i}}
\newcommand\rmK{\mathrm{K}}
\newcommand\rmm{\mathrm{m}}

\newcommand\rms{\mathrm{s}}
\newcommand\rmT{\mathrm{T}}
\newcommand\f{\frac}
\newcommand\p{\partial}

\newcommand\he{\mathrm{He}}

\title[Three-dimensional eccentric discs around Be stars]
{Three-dimensional eccentric discs around Be stars}

\author[Gordon I. Ogilvie]{Gordon I. Ogilvie\\
Department of Applied Mathematics and Theoretical Physics,
University of Cambridge, Centre for Mathematical Sciences,\\
Wilberforce Road, Cambridge CB3 0WA
}

\begin{document}

\maketitle

\label{firstpage}
 
\begin{abstract}
  One-armed oscillation modes in the circumstellar discs of Be~stars
  may explain the cyclical variations in their emission lines.  We
  show that a three-dimensional effect, involving vertical
  motion and neglected in previous treatments, profoundly influences
  the dynamics.  Using a secular theory of eccentric discs that
  reduces the problem to a second-order differential equation, we show
  that confined prograde modes are obtained for all reasonable disc
  temperatures and stellar rotation rates.  We confirm these results
  using a numerical analysis of the full set of linearized equations
  for three-dimensional isothermal discs including viscous
  terms that couple the horizontal motions at different altitudes.  In
  order to make these modes grow, viscous damping must be overcome by
  an excitation mechanism such as viscous overstability.
\end{abstract}


\section{Introduction}

Classical Be~stars \citep{2003PASP..115.1153P} are rapidly rotating
early-type stars that exhibit Balmer emission lines.  It is widely
agreed that these lines, which are generally double-peaked, originate
in a relatively thin circumstellar disc that is in approximately
Keplerian rotation \citep[and references
therein]{2007ASPC..361..230O}.  While the precise mechanism by which
the disc is formed remains controversial, it is likely to resemble a
viscous decretion disc that is expelled by the action of a torque at
its inner boundary \citep{1991MNRAS.250..432L,2003PASP..115.1153P}.

Many Be~stars show cyclical variations in their double-peaked
emission lines over years or decades, with the red and blue
peaks alternately becoming more prominent \citep[and references
therein]{1997A&A...318..548O}.  An explanation of this phenomenon was
given by \citet{1991PASJ...43...75O}, who proposed that a
low-frequency, one-armed oscillation mode \citep{1983PASJ...35..249K}
occurs in the disc.  This is equivalently to saying that the disc
becomes eccentric and the slow precession of its elliptical shape
gives rise to the cyclical changes in the observed emission lines.

Eccentric discs, in which fluid elements, solid particles or stars
follow elliptical orbits of variable eccentricity around a central
mass, have further applications in systems as diverse as planetary
rings, protoplanetary systems, close binary stars and galactic nuclei.
A detailed understanding of the origin of eccentricity and the rates
of precession in the circumstellar discs of Be~stars would therefore
be of general interest.

\citet{1991PASJ...43...75O} originally considered a disc that orbits
in a point-mass potential and obtained a sequence of retrograde modes
in which the precession of the disc is in a direction opposite to its
rotation.  Retrograde precession is a natural consequence of the
pressure forces in the disc, which cause a small departure from
Keplerian rotation and allow the eccentricity to propagate in a
wavelike manner.  The global modes found by
\citet{1991PASJ...43...75O} are weighted towards the outer part of the
disc and their periods become extremely long as the outer radius of
the disc is increased to realistic values.

\citet{1992A&A...265L..45P} and \citet{1993A&A...276..409S} considered
the effect of the quadrupole gravitational potential associated with
the rotational deformation of the star, which tends to cause a
prograde precession of elliptical orbits.  They showed that, when the
quadrupole effect is taken into account, prograde modes can be
obtained that are naturally confined in the inner part of the disc and
are insensitive to the outer boundary condition.  Subsequent
observations confirmed that the precession is indeed prograde
\citep{1994A&A...288..558T}.

However, \citet{1997A&A...318..548O} found that confined prograde
modes can be obtained only when the disc is sufficiently cool, so that
the quadrupole effect dominates over the tendency of pressure to
produce retrograde precession and extended modes.  He concluded that a
different mechanism is required in the hotter discs of early-type
Be~stars such as the prototype $\gamma$~Cas, and proposed
that radiative line forces could explain the required prograde
precession.  Unfortunately the modelling of radiative forces is
subject to considerable uncertainty.  \citet{2006A&A...447..277F}
found that the resulting model has little predictive power owing to
the sensitivity of the results to the parameters.

More recently, \citet{2006A&A...456.1097P} investigated an alternative
way to obtain prograde modes in hotter discs.  This involves replacing
the rigid inner boundary condition at the stellar surface with a free
boundary condition, on the basis that a gap is formed between the star
and the disc.  Such a gap is not generally expected in the scenario of
the viscous decretion disc but might be possible in alternative
models.

All of the treatments described so far are based on two-dimensional
models of the disc that neglect aspects of its vertical structure and
motion.  At first sight, such an approach seems reasonable for
studying eccentric modes in thin discs, where the motion might be
assumed to be purely horizontal and independent of height.  However,
in presenting a three-dimensional, non-linear theory of eccentric
discs, we have previously argued that three-dimensional effects are of
considerable importance \citep{2001MNRAS.325..231O}.  In particular,
the variation of the vertical gravitational acceleration around an
elliptical orbit excites an oscillatory vertical motion in an
eccentric disc that should not be neglected.

In this paper we show that, in fact, three-dimensional effects are
essential to understanding the precession of eccentric discs around
Be~stars.  They allow confined prograde modes to be obtained even when
the stellar quadrupole moment is negligible and when the inner
boundary is rigid.  This property allows us to give a unified
description of eccentric discs of Be~stars of all stellar types
without introducing uncertain radiative forces or modifying the inner
boundary condition.  While we do not deny that in some cases radiative
forces might be important, or that the inner boundary condition might
differ from a rigid one, we show that these innovations are
unnecessary.

Most previous analyses have not discussed the processes that
could cause eccentric modes to grow or decay, and which are therefore
relevant to explaining the occurrence of eccentric discs around
Be~stars.  Viscous overstability
\citep{1978MNRAS.185..629K,2006MNRAS.372.1829L} provides a possible
explanation; \citet{2001A&A...369..117N} have applied this idea to
Be~stars and estimated the associated growth rate.  We defer to a
future investigation a detailed analysis of the effects of viscous or
turbulent stresses.  A non-linear treatment, which could address the
saturation of the growth mechanism and attempt to predict the observed
amplitudes and precession rates of the eccentric modes, also remains
to be undertaken.

The structure of this paper is as follows.  In Section~\ref{s:basic}
we describe the basic state of the disc.  We then discuss an
approximate, secular theory of three-dimensional eccentric discs in
Section~\ref{s:secular}, comparing it with the two-dimensional
theories used by other authors.  In Section~\ref{s:full} we solve the
full linearized equations accurately using a spectral method,
including some effects of viscosity, and compare the results with
those of the secular theory.  Conclusions are given in
Section~\ref{s:conc}.

\section{Basic state of the disc}
\label{s:basic}

We consider a basic state consisting of a steady, axisymmetric disc
around a rotating star.  Adopting cylindrical polar coordinates
$(r,\phi,z)$, we write the gravitational potential of the star as
\begin{equation}
  \Phi=-GM(r^2+z^2)^{-1/2}\left[1+
  \f{Q}{3}\f{(r^2-2z^2)R^2}{(r^2+z^2)^2}\right],
\end{equation}
where $M$ and $R$ are its mass and (equatorial) radius.  This
potential consists of monopole and quadrupole components, the latter
being of dimensionless strength $Q$ and arising from the rotational
deformation of the star.  (The parameter $Q$ is related to the
gravitational moment $J_2$ used in planetary science by $Q=3J_2/2$.)
For a star with uniform angular velocity $\Omega_*$, we have
$Q=k_2\Omega_*^2R^3/GM$, where $k_2$ is the apsidal motion constant.
We neglect higher-order multipole components as well as the
self-gravitation of the disc.

In common with most other treatments, we assume that the stellar
radiation maintains the disc at a constant temperature $T_\rmd$,
slightly less than the effective temperature $T_\rme$ of the star.
The isothermal sound speed $c_\rms=(\mathcal{R}T_\rmd/\mu)^{1/2}$ is
therefore constant, both in the equilibrium state and for the
perturbations.  We neglect any viscous or turbulent stresses and any
resulting meridional motions in the disc.

The basic state then has velocity $\bu=r\Omega(r)\,\be_\phi$.  The
angular velocity $\Omega$ is independent of $z$ because the basic
state is isothermal and therefore barotropic.  The balance of forces
requires $r\Omega^2\,\be_r=\bnabla(\Phi+h)$, where $h=c_\rms^2\ln\rho$
is the (pseudo-)enthalpy of an isothermal gas.

The potential and enthalpy in the midplane are
$\Phi_\rmm(r)=\Phi(r,0)=-(GM/r)[1+(Q/3)(R/r)^2]$ and
$h_\rmm(r)=h(r,0)$, and the radial force balance implies
\begin{equation}
  \Omega^2=\f{GM}{r^3}\left[1+Q\left(\f{R}{r}\right)^2\right]+\f{1}{r}\p_rh_\rmm.
\label{omsq}
\end{equation}
The associated epicyclic frequency $\kappa(r)$ is given by
\begin{eqnarray}
  \kappa^2&=&\f{1}{r^3}\p_r(r^4\Omega^2)\nonumber\\
  &=&\f{GM}{r^3}\left[1-Q\left(\f{R}{r}\right)^2\right]+\f{1}{r^3}\p_r(r^3\p_rh_\rmm).
\label{kasq}
\end{eqnarray}
We also refer to the Keplerian angular velocity $\Omega_\rmK(r)$
defined by
\begin{equation}
  \Omega_\rmK^2=\f{GM}{r^3}.
\end{equation}

For a thin disc we may expand the potential about the midplane to
obtain $\Phi\approx\Phi_\rmm+\half\Omega_z^2z^2$, where the vertical
frequency $\Omega_z(r)$ is given by
\begin{equation}
  \Omega_z^2=\f{GM}{r^3}\left[1+3Q\left(\f{R}{r}\right)^2\right].
\end{equation}
The vertical force balance then implies $h=h_\rmm-\half\Omega_z^2z^2$
and so $\rho=\rho_\rmm\exp[-(z^2/2H^2)]$, where $\rho_\rmm(r)$ is the
midplane density and $H(r)=c_\rms/\Omega_z$ is the scaleheight.  The
surface density is $\Sigma(r)=(2\pi)^{1/2}\rho_\rmm H$.

Since the basic state is steady and axisymmetric, wave modes may be
considered in which the dependence on azimuth and time is of the form
$\exp(\rmi m\phi-\rmi\omega t)$, where $m$ is the azimuthal wavenumber
and $\omega$ is the wave frequency.  In the case $m=1$ that is of
interest here, $\omega$ is also the angular pattern speed of the mode
in an inertial frame of reference.  This can be identified with the
precession rate of the eccentric disc, which is positive if the
precession is prograde (i.e.\ in the same direction as the rotation of
the disc).

\section{Secular treatment}
\label{s:secular}

\subsection{Comparison of two- and three-dimensional theories}

An approximate theory can be derived in which the disc is assumed to
be nearly Keplerian and the frequency $\omega$ of the wave is much
less than the orbital frequency $\Omega$ of the disc.  This type of
approximation, which is related to the secular theory of celestial
mechanics, has been discussed in the two-dimensional linear case by
\citet{2001AJ....121.1776T}, \citet{2002A&A...388..615P} and
\citet{2006MNRAS.368.1123G}, and in the three-dimensional non-linear
case by \citet{2001MNRAS.325..231O}.  The derivation of this theory in
the case of a three-dimensional isothermal disc around a Be~star is
presented in Appendix~\ref{s:appendix}, which draws on the analysis of
Section~\ref{s:full} below.  The two-dimensional approximation is also
discussed there.

In the secular theory the radial velocity $u(r)$ satisfies the
equation
\begin{equation}
  (\omega-f)u=\f{1}{2r^2}\p_r\left[\f{c_\rms^2r^3}{\Sigma}\p_r\left(\f{\Sigma u}{r\Omega}\right)\right],
\label{secular}
\end{equation}
where $f(r)$ is given by
\begin{equation}
  f=\f{\Omega^2-\kappa^2}{2\Omega}
\label{f2d}
\end{equation}
in a two-dimensional disc, but
\begin{equation}
  f=\f{\Omega^2-\kappa^2}{2\Omega}+\f{9c_\rms^2}{4r^2\Omega}
\label{f3d}
\end{equation}
in a three-dimensional disc.  With the dependence
$\rme^{\rmi\phi-\rmi\omega t}$ assumed above, the radial velocity is
related to the complex eccentricity $E(r)=e\,\rme^{\rmi\varpi}$, where
$e(r)$ is the eccentricity and $\varpi(r)$ is the longitude of
periastron measured in a frame of reference that rotates with the
angular pattern speed $\omega$ of the mode, by $u^*=\rmi r\Omega E$
\citep{2001MNRAS.325..231O}.  It can be seen from
equation~(\ref{secular}) that $f$ is a local contribution to the
global precession rate $\omega$ of the eccentric mode.  Indeed, an
integral expression for the mode frequency in the case of rigid
boundary conditions\footnote{Although there is no physical
justification for a rigid outer boundary condition, a confined mode
decays sufficiently fast with increasing $r$ that the same integral
relation is recovered in the limit of large $r_\mathrm{out}$.  Such
modes are in fact insensitive to the outer boundary condition.} ($u=0$
at $r=r_\mathrm{in}$ and $r=r_\mathrm{out}$) is
\begin{eqnarray}
  \lefteqn{\omega\int_{r_\mathrm{in}}^{r_\mathrm{out}}\f{\Sigma r|u|^2}{\Omega}\,\rmd r=\int_{r_\mathrm{in}}^{r_\mathrm{out}}f\,\f{\Sigma r|u|^2}{\Omega}\,\rmd r}&\nonumber\\
  &&-\int_{r_\mathrm{in}}^{r_\mathrm{out}}\f{c_\rms^2r^3}{2\Sigma}\left|\p_r\left(\f{\Sigma u}{r\Omega}\right)\right|^2\,\rmd r,
\label{vp}
\end{eqnarray}
which follows from equation~(\ref{secular}) after multiplication by
$\Sigma ru^*/\Omega$ and an integration by parts.  The first term on
the right-hand side shows the contribution to the precession rate from
$f$ in the form of an integral weighted by the structure of the mode,
while the second term shows a retrograde contribution associated with
pressure.  (Note that the pressure also contributes indirectly through
its effect on $f$.)

In a two-dimensional disc $f$ is given by equation~(\ref{f2d}) and
corresponds to the expected expression for the local apsidal
precession rate.  Note that, in this case, $f\approx\Omega-\kappa$ if,
as assumed, the disc is nearly Keplerian
($|\Omega^2-\kappa^2|\ll\Omega^2$).  There is, in fact, more than one
version of the two-dimensional theory.  \citet{1993A&A...276..409S}
work throughout with vertically averaged equations and find
\begin{equation}
  f=Q\f{GMR^2}{r^5\Omega}-\f{c_\rms^2}{2r^2\Omega}\p_r(r^2\p_r\ln\Sigma).
\label{2ds}
\end{equation}
\citet{1997A&A...318..548O} calculates $\Omega$ and $\kappa$ using a
three-dimensional equilibrium disc but then applies vertically
averaged equations for the perturbations.  He therefore uses
\begin{equation}
  f=Q\f{GMR^2}{r^5\Omega}-\f{c_\rms^2}{2r^2\Omega}\p_r(r^2\p_r\ln\rho_\rmm),
\label{2do}
\end{equation}
which agrees with equations~(\ref{omsq}) and~(\ref{kasq}) above.

In a three-dimensional disc, however, equation~(\ref{f3d}) gives the
relevant expression for $f$ as
\begin{equation}
  f=Q\f{GMR^2}{r^5\Omega}-\f{c_\rms^2}{2r^2\Omega}\p_r(r^2\p_r\ln\rho_\rmm)+
  \f{9c_\rms^2}{4r^2\Omega}.
\label{3d}
\end{equation}
The last term represents an additional local contribution to the
prograde precession of the disc that arises from the three-dimensional
dynamics including the vertical motion; it is discussed further in
Section~\ref{s:interp} below.  For the parameters relevant to
Be~stars, this term is never negligible (it corresponds to a
precession period of order $1\,\mathrm{yr}$ at the stellar surface)
and declines much more slowly with radius than the quadrupole term.

We refer to the three theories described above as `2DS'
\citep[equation~\ref{2ds}
above;][]{1993A&A...276..409S,2006A&A...456.1097P}, `2DO'
\citep[equation~\ref{2do}
above;][]{1991PASJ...43...75O,1997A&A...318..548O} and `3D'
(equation~\ref{3d}).  Note, however, that the secular approximations
were usually not employed in the cited papers.  The 3D~secular theory
is based on similar assumptions and approximations to the non-linear
analysis of \citet{2001MNRAS.325..231O}.  In particular, some viscous
or other stress is required to couple different layers in the disc so
that they tend to adopt the same eccentricity.  In the absence of such
stresses the eccentricity may vary significantly with $z$ and the
results can differ.  This complication is discussed in
Section~\ref{s:full}, where the full linearized equations are solved.

\subsection{Application to power-law discs}
\label{s:powerlaw}

It is consistent with the spirit of the secular approximation to
neglect the differences between $\Omega$, $\kappa$, $\Omega_z$ and
$\Omega_\rmK$ except where essential (i.e.\ in the quantity
$\Omega^2-\kappa^2$).  Accordingly, we may take $\Omega\propto
r^{-3/2}$ and $H\propto r^{3/2}$.  For a surface density profile
$\Sigma\propto r^{-\sigma}$ the density in the midplane varies as
$\rho_\rmm\propto r^{-\sigma-3/2}$.  We introduce the dimensionless
radial coordinate
\begin{equation}
  x=\f{r}{R}
\end{equation} 
and the small parameter
\begin{equation}
  \epsilon=c_\rms\left(\f{R}{GM}\right)^{1/2},
\end{equation}
which is a measure of the angular semithickness $H/r$ of the disc at the
stellar surface.  The three theories then give
\begin{equation}
  \mathrm{2DS:\phantom{O3D}}\f{f}{\Omega}=Qx^{-2}+\half\sigma\epsilon^2x,
\end{equation}
\begin{equation}
  \mathrm{2DO:\phantom{S3D}}\f{f}{\Omega}=Qx^{-2}+\half(\sigma+{\textstyle\f{3}{2}})\epsilon^2x,
\end{equation}
\begin{equation}
  \mathrm{3D:\phantom{S2DO}}\f{f}{\Omega}=Qx^{-2}+\half(\sigma+6)\epsilon^2x.
\end{equation}
We therefore consider the generic form
\begin{equation}
  \f{f}{\Omega}=Qx^{-2}+s\epsilon^2x,
\end{equation}
where $s$ is a constant.  It is convenient to work with dimensionless,
rescaled values of the mode frequency and quadrupole strength,
$\tilde\omega$ and $\tilde Q$, defined by
\begin{equation}
  \omega=\epsilon^2\left(\f{GM}{R^3}\right)^{1/2}\tilde\omega,
\end{equation}
\begin{equation}
  Q=\epsilon^2\tilde Q.
\end{equation}
Equation~(\ref{secular}) then becomes
\begin{eqnarray}
  \lefteqn{\left[\tilde\omega-(\tilde Qx^{-7/2}+sx^{-1/2})\right]u}&\nonumber\\
  &&=\f{1}{2x^2}\p_x\left[x^{\sigma+3}\p_x(x^{-\sigma+1/2}u)\right],
\label{secular2}
\end{eqnarray}
which is an eigenvalue problem of Sturm--Liouville form, when
considered together with appropriate boundary conditions.  Note that
the parameter $\epsilon$ drops out of the equation under the above
rescalings.  For prograde modes ($\tilde\omega>0$), one solution of
this equation decays exponentially as $x\to\infty$, while the other
solution grows exponentially.  We therefore consider the equation in
the domain $1<x<\infty$, requiring solutions to satisfy a rigid
boundary condition $u=0$ at the stellar surface $x=1$ and to decay as
$x\to\infty$.

Reasonable values of $\tilde Q$ for Be~stars can be estimated
following \citet{2006A&A...447..277F}.  Their Table~4 is based on
stellar models of intermediate main-sequence age.  There is
considerable uncertainty in the applicability of these models because
of the rapid rotation of Be~stars.  Nevertheless, using these data, we
estimate that $\tilde Q$ ranges from approximately~$8$ at the lower
end ($M=2.51\,M_\odot$) to approximately~12 at the upper end
($M=15.85\,M_\odot$).  These values assume, somewhat arbitrarily, a
disc temperature of $T_\rmd=(2/3)T_\rme$ and a stellar rotation rate
that is $95\%$ of the critical value.  Similarly, for the stellar
models of $\gamma$~Cas and $59$~Cyg described in Tables~2 and~3 of
\citet{2006A&A...447..277F}, we estimate $\tilde Q\approx9$ and~$13$,
respectively.  It is clear that significantly larger values of $\tilde
Q$ cannot be obtained by further increasing the stellar rotation rate,
and in fact smaller values may be more appropriate.  The values of
$\epsilon$ for all these models are in the range $0.021$--$0.026$, again
assuming that $T_\rmd=(2/3)T_\rme$.

\subsection{Solutions in the absence of a quadrupole}

Consider first the case without a quadrupole term, $\tilde Q=0$.  The
solution that decays as $x\to\infty$ is then
\begin{equation}
  u=x^{(\sigma-3)/2}K_\nu(z),
\label{knu}
\end{equation}
with
\begin{equation}
  \nu=2[(\sigma+2)^2-8s]^{1/2},\qquad z=4(2\tilde\omega)^{1/2}x^{1/4},
\end{equation}
where $K_\nu$ is the modified Bessel function of the second kind of
order $\nu$ \citep{1965hmfw.book.....A}.  Note that $z$ is real and
positive for the prograde modes of interest.  This solution has no
zeros in $z>0$ if $\nu$ is real, but does so if $\nu$ is imaginary.
To match a rigid boundary condition at $x=1$ we therefore require
$\nu^2<0$.  The 2DS~theory gives $8s=4\sigma$, so
$\nu^2=4(\sigma^2+4)$ and modes are never confined.  The 2DO~theory
gives $8s=4\sigma+6$, so $\nu^2=4(\sigma^2-2)$ and modes may be
confined for $\sigma^2<2$.  The 3D~theory gives $8s=4\sigma+24$, so
$\nu^2=4(\sigma^2-20)$ and modes may be confined for $\sigma^2<20$.

\begin{figure}
\centerline{\epsfysize8cm\epsfbox{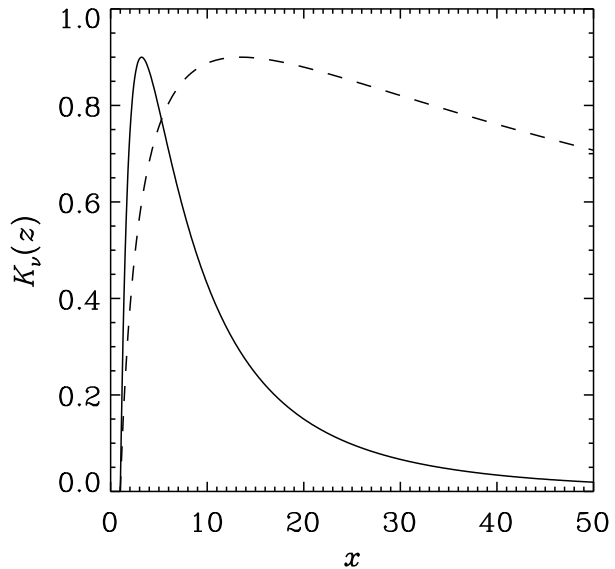}}
\caption{Confinement of prograde modes in the absence of a quadrupole.
The modified Bessel function $K_\nu(z)$ (multiplied by a constant for
convenient normalization) is plotted against $x=r/R$ for the cases
$\nu=8\rmi$ (solid line) and $\nu=\sqrt{8}\rmi$ (dashed line).
According to equation~(\ref{knu}), the modified Bessel function, which
decays exponentially as $x\to\infty$ if $\nu$ is imaginary, describes
the confinement of the eigenfunction.  The solid line is relevant to
the 3D~theory for a surface density index $\sigma=2$, and shows a good
confinement of the mode.  The dashed line is relevant to the
2DO~theory in the (optimal) case $\sigma=0$, but here the confinement
is very poor.}
\label{f:knu}
\end{figure}

This analysis is somewhat misleading because the modes may not be
adequately confined to be applicable to the discs of Be~stars.  In
Fig.~\ref{f:knu} we plot $K_\nu(z)$ versus $x$ for $\nu=8\rmi$ and
$\nu=\sqrt{8}\rmi$, choosing $\tilde\omega$ such that $K_\nu(z)=0$ at
$x=1$.  Only the former case provides an adequately confined mode.
This shows that the 2DO~theory cannot in fact produce confined
prograde modes in the absence of a quadrupole, even for the most
optimistic choice of surface density profile, because $|\nu|$ is never
large enough.  However, sufficiently large imaginary values of $\nu$
are obtained in the 3D~theory.

When $\nu$ is imaginary, confined solutions are obtained, in
principle, when $\tilde\omega=z_n^2/32$ where $z_{n,\nu}$ is the $n$th
zero of $K_\nu(z)$ in $z>0$.  For example, consider the 3D~theory with
$\sigma=2$, which is the value expected for a steady decretion disc
with constant alpha viscosity parameter, far inside its outer radius.
Then $s=4$ and $\nu=8\rmi$, and zeros occur at $z=4.802$, $3.067$,
$2.029$, $\dots$\ Therefore, in principle, $\tilde\omega=0.7207$,
$0.2940$, $0.1286$, $\dots$ are the scaled frequencies of confined
modes.  However, only the first of these, corresponding to an
eigenfunction with no nodes (Fig.~\ref{f:knu}, solid line), gives rise
to a mode that is adequately confined in a disc of modest radial
extent.

\subsection{Critical quadrupole strength}
\label{s:qcrit}

The integral expression for the dimensionless frequency eigenvalue of
a mode that satisfies a rigid boundary condition at the stellar
surface and decays as $r\to\infty$ is (cf.\ equation~\ref{vp})
\begin{eqnarray}
  \lefteqn{\tilde\omega\int_1^\infty x^{-\sigma+5/2}u^2\,\rmd x=
  \int_1^\infty(\tilde Qx^{-\sigma-1}+sx^{-\sigma+2})u^2\,\rmd x}&\nonumber\\
  &&-\f{1}{2}\int_1^\infty x^{\sigma+3}\left[\p_x(x^{-\sigma+1/2}u)\right]^2\,\rmd x,
\end{eqnarray}
where we take $u$ to be real.  Following \citet{2006A&A...456.1097P},
we note that this expression has the usual variational property
associated with self-adjoint eigenvalue problems.  In this case, a
confined prograde mode exists if and only the right-hand side of this
equation can be made positive by a trial function $u(x)$ satisfying
the appropriate boundary conditions.  This is clearly possible if
either $\tilde Q$ or $s$ is large enough.  What is the minimum value
of $\tilde Q$ (for a given value of $s$) such that the right-hand side
can just be made to vanish for a non-trivial $u$?  This happens when
\begin{eqnarray}
  \lefteqn{\tilde Q\int_1^\infty x^{-\sigma-1}u^2\,\rmd x=
  \f{1}{2}\int_1^\infty
  x^{\sigma+3}\left[\p_x(x^{-\sigma+1/2}u)\right]^2\,\rmd
  x}&\nonumber\\ &&-\int_1^\infty sx^{-\sigma+2}u^2\,\rmd x.
\end{eqnarray}
The minimum value of $\tilde Q$ for which this equation can be
satisfied is given by the corresponding Euler--Lagrange equation,
which is just equation~(\ref{secular2}) with $\tilde\omega$ set to
zero.  The solutions are
\begin{equation}
  u=x^{(\sigma-3)/2}J_{\pm\mu}(w),
\end{equation}
with
\begin{equation}
  \mu=\f{\nu}{6}=\f{1}{3}[(\sigma+2)^2-8s]^{1/2},\qquad w=\f{\tilde Q^{1/2}}{3}\left(\f{x}{2}\right)^{-3/2}.
\end{equation}
If $\mu^2<0$ we have seen that confined modes can be found, in
principle, even if $\tilde Q=0$.  Consider then the case $\mu^2>0$.
The solution that decays\footnote{Since $J_{\mu}(w)\propto w^\mu$ for
small $w$, $u\propto x^{[\sigma-3(1+\mu)]/2}$ for large $x$.  This
solution satisfies the condition that
$x^{7/2}u\p_x(x^{-\sigma+1/2}u)\to0$ as $x\to\infty$, which is
required to carry out the integration by parts and derive the
variational principle and Euler--Lagrange equation, while $J_{-\mu}$
does not.} as $x\to\infty$ is $J_{\mu}(w)$.  The critical condition
for a rigid inner boundary condition is therefore $\tilde
Q=9w_\mu^2/8$, where $w_\mu$ is the first zero of $J_\mu(w)$ in $w>0$.
(This is an increasing function of $\mu$ and therefore a decreasing
function of $s$.)

For example, if $\sigma=2$, we require $\tilde Q>10.8$ for confinement
in the 2DO~theory, or $\tilde Q>15.9$ in the 2DS~theory.  These values
are not very sensitive to $\sigma$; \citet{2006A&A...456.1097P} quote
$\tilde Q>17.3$ for $\sigma=5/2$ and a rigid inner boundary condition,
which agrees with this analysis.

\subsection{Schr\"odinger analogy}

\begin{figure}
\centerline{\epsfysize8cm\epsfbox{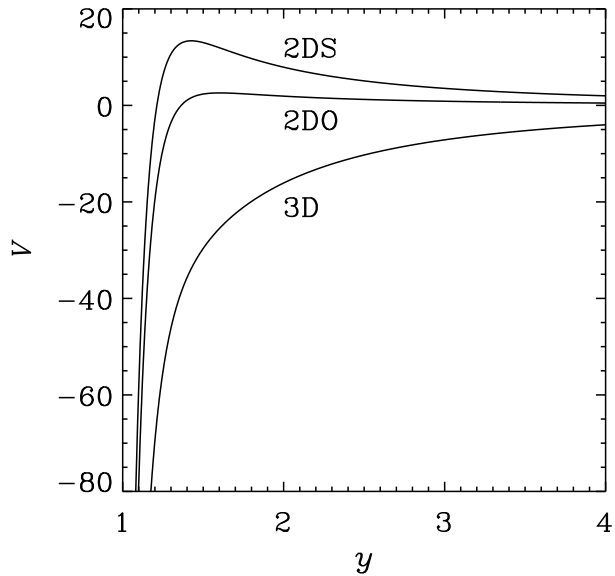}}
\caption{Effective potentials $V(y)$ (equation~\ref{v}) describing the
confinement of prograde modes by analogy with bound states in quantum
mechanics.  Potentials are shown for the three theories for the case
$\sigma=2$ and $\tilde Q=10$.  The bottom of the potential well at
$y=1$ (i.e.\ at the stellar surface) in each case is considerably
deeper than can be shown on the graph.  However, only in the 3D~theory
is the well deep and wide enough in this case to support a bound state
in the case of a rigid inner boundary condition.}
\label{f:v}
\end{figure}

A useful description of confined modes uses an analogy with bound
states in quantum mechanics.  If $y=x^{1/4}$ and
$u=x^{\sigma/2-13/8}\psi(y)$, then $\psi(y)$ satisfies the
Schr\"odinger equation
\begin{equation}
  -\f{\rmd^2\psi}{\rmd y^2}+[V(y)-E]\psi=0
\end{equation}
with an effective potential
\begin{equation}
  V=\f{1}{4}(16\sigma^2+64\sigma+63-128s)y^{-2}-32\tilde Qy^{-14}
\label{v}
\end{equation}
and an effective energy eigenvalue $E=-32\tilde\omega$.  The
coefficient of $1/4y^2$ in $V$ is $(16\sigma^2+63)$ for the
2DS~theory, $(16\sigma^2-33)$ for 2DO and $(16\sigma^2-321)$ for 3D.
It is therefore likely to be negative only in the 3D~theory.  Note
that the disc occupies the region $y>1$.  A bound state of negative
energy, equivalent to a confined prograde mode, can be obtained if
there is a sufficiently deep and wide potential well.  The $\tilde Q$
term tends to create a deep well close to the stellar surface.  In the
2D~theories it competes with the $y^{-2}$ term, which contributes a
repulsive potential.  In the 3D~theory, however, the $y^{-2}$ term
creates a much wider well, so allowing broader modes with slower
precession rates, even if $\tilde Q=0$.  These potentials are plotted
in Fig.~\ref{f:v} for the case $\sigma=2$ and $\tilde Q=10$.  For
these parameters the 3D~potential is deep and wide enough to support a
bound state in the case of a rigid inner boundary condition,
while the others are not.

\subsection{Precession rates and mode shapes}

\begin{figure*}
\centerline{\epsfysize8.5cm\epsfbox{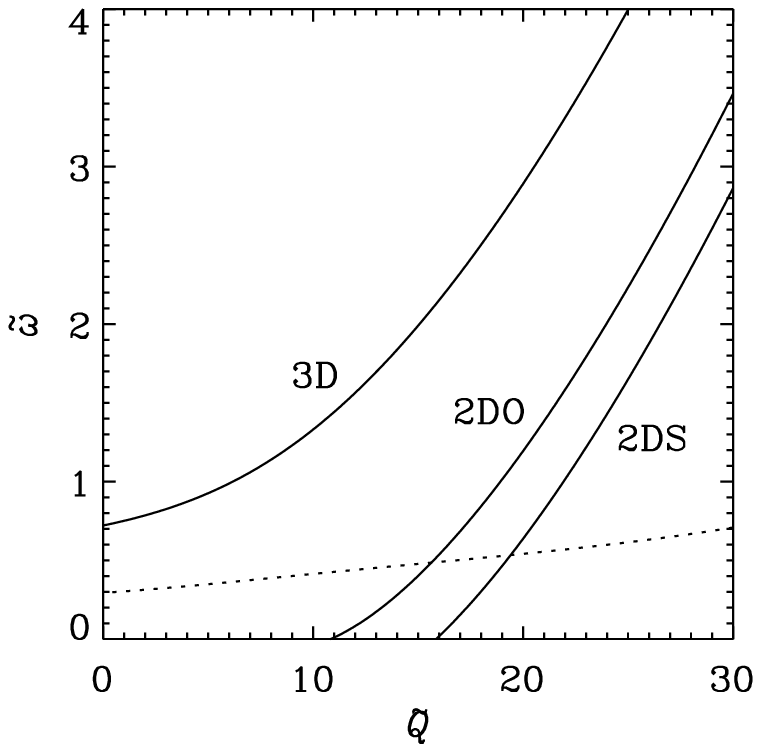}\qquad\epsfysize8.5cm\epsfbox{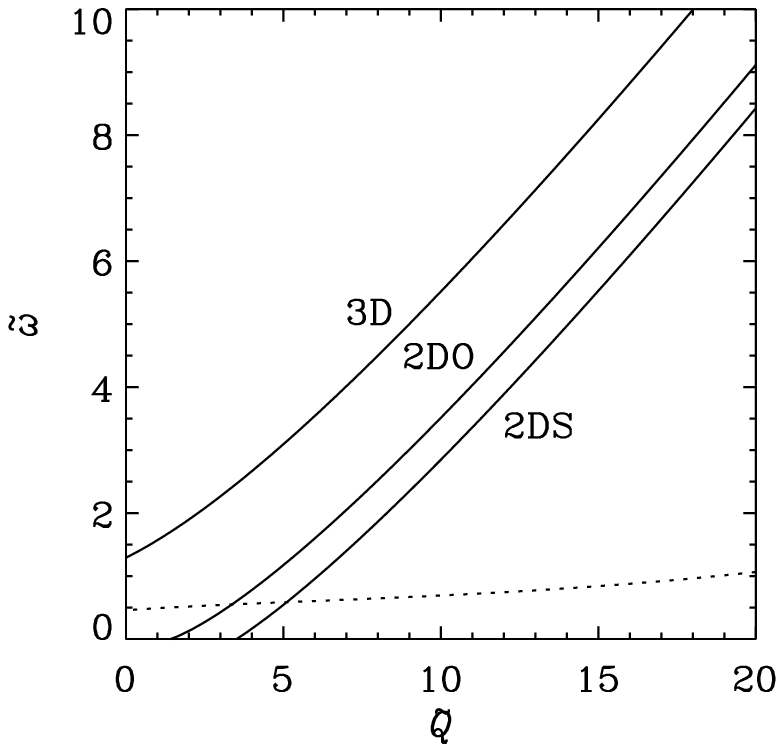}}
\caption{Dimensionless precession rates of confined prograde modes in
the three theories, for the case $\sigma=2$.  The left and
right panels are for rigid and free inner boundary conditions,
respectively.  These results are insensitive to the outer boundary
condition provided that the disc is reasonably large.  In the
2D~theories, confined prograde modes are found only for sufficiently
large values of $\tilde Q$.  The dotted line indicates a
second confined mode, with one radial node in its eigenfunction, found
in the 3D~theory; this and higher-order modes are less well confined
and therefore more sensitive to the outer boundary condition.}
\label{f:ot}
\end{figure*}

We now compute the eigenvalues $\tilde\omega$ of
equation~(\ref{secular2}) by a shooting method, adopting a rigid inner
boundary condition at the stellar surface and seeking prograde
confined modes that decay as $x\to\infty$.  (In practice this is done
by imposing a rigid outer boundary condition and verifying that the
eigenvalue is completely insensitive to the value of the outer radius
provided it is sufficiently large.)  The results are shown in the left
panel of Fig.~\ref{f:ot}.  Here we confirm that confined prograde
modes exist in the 2D~theories only for sufficiently large values of
$\tilde Q$, as described in Section~\ref{s:qcrit}.  In contrast, the
3D~theory allows such modes to be obtained for any value of $\tilde
Q$.  Furthermore, the precession rates are much larger in the
3D~theory.  Since the realistic values of $\tilde Q$ are probably in
the vicinity of $10$ or smaller, it is clear that the
three-dimensional effects are of essential importance in the
case of a rigid inner boundary condition.

In the right panel of Fig.~\ref{f:ot} we show comparable
results for the free inner boundary condition considered by
\citet{2006A&A...456.1097P}, meaning that the Lagrangian pressure
perturbation is zero at $r=R$.  In the secular theory this is
equivalent to $\p_rE=0$, or $u+2r\p_r u=0$.  As described by
\citet{2006A&A...456.1097P}, a free inner boundary condition allows
prograde modes to be found for smaller values of $\tilde Q$ in the
2DS~theory.  It is still true, however, that the three-dimensional
effects have an important effect on the results.  Since the
eigenfunctions obtained with a free inner boundary condition are
generally peaked at the inner radius, there may also be a conflict
between the nonlinear development of such an eccentric mode and the
existence of a stellar surface, unless there is a wide gap between the
star and the disc.

To convert these eigenvalues into physical units, we again make use of
the stellar models in \citet{2006A&A...447..277F}.  We then find that
the precession period is
\begin{equation}
  P=\f{C_P}{\tilde\omega}\,\mathrm{yr},
\end{equation}
where $C_P$ ranges from approximately $1.6$ at the lower end to
approximately $2.3$ at the upper end (or $2.4$ for $\gamma$~Cas and
$1.8$ for $59$~Cyg).  This conversion factor is independent of
assumptions regarding the stellar rotation rate, except inasmuch as
the rotation affects the equatorial radius, but $C_P$ is inversely
proportional to the assumed disc temperature.

\begin{figure}
\centerline{\epsfysize8cm\epsfbox{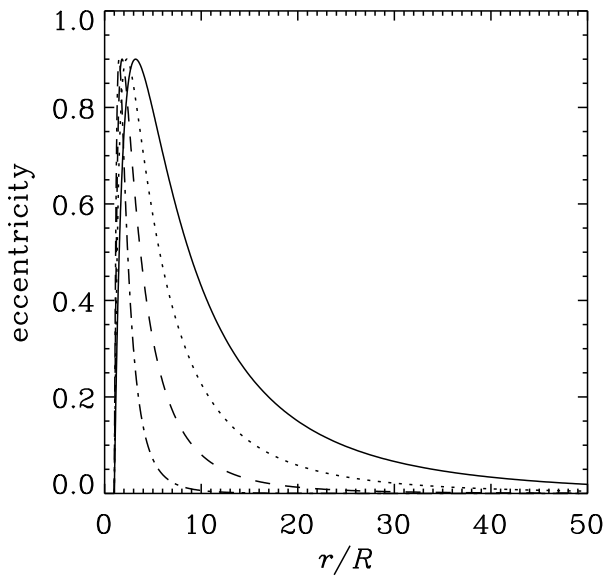}}
\caption{Eigenfunctions of confined prograde modes in the 3D~secular
theory with a rigid inner boundary condition.  The eccentricity
(proportional to $u/r\Omega$, and with arbitrary normalization) is
plotted versus radius for the case $\sigma=2$, and for relative
quadrupole strengths $\tilde Q=0$ (solid line), $5$ (dotted line),
$10$ (dashed line) and $20$ (dot-dashed line).  The first case
corresponds to the solid line in Fig.~\ref{f:knu}.}
\label{f:ef}
\end{figure}

The shapes of the confined modes in the 3D~theory are illustrated in
Fig.~\ref{f:ef}.  For larger values of $\tilde Q$, the mode is
increasingly confined in the inner part of the disc.  Although
higher-order modes may exist, one of which is referred to in
Fig.~\ref{f:ot}, we focus here on the fundamental confined mode with
the simplest radial structure.

Since $\tilde\omega$ is close to~$1$ in the 3D~theory for reasonable
values of~$\tilde Q$ and with a rigid inner boundary
condition, precession rates in the vicinity of $1$--$3\,\mathrm{yr}$
are obtained.  Observed cycle times, which can vary significantly even
for the same star, are more usually in the range
$5$--$10\,\mathrm{yr}$ \citep{1997A&A...318..548O}.  This discrepancy
might occur because the discs have different temperature or density
profiles from those assumed here, either of which would affect the
precession period.  Non-linearity of the eccentric mode may also
increase the precession period by altering the distribution of
eccentricity so that it is less peaked in the inner part of the disc.
Furthermore, non-isothermal effects or radiative forces might be
important.  In some cases, including $\gamma$~Cas and $59$~Cyg, the
presence of a relative close binary companion may also affect the
dynamics of the eccentric mode, although it would seem most
likely to contribute to the prograde precession and therefore to
decrease the precession period.

\subsection{Physical interpretation of the three-dimensional dynamics}
\label{s:interp}

What is the origin of the additional prograde precession that occurs
in a three-dimensional disc?  As described in
Appendix~\ref{s:appendix} and Section~\ref{s:full} below, two
different types of motion are involved in an eccentric disc.  One
(corresponding to the $n=0$ mode in later sections) consists of
horizontal velocities and enthalpy perturbations that are independent
of $z$; this describes the eccentric orbital motion of the gas.  The
other (corresponding to $n=2$) involves a vertical velocity
proportional to $z$ and an enthalpy perturbation proportional to
$z^2-H^2$; this is a vertical `breathing' mode of the disc.  Coupling
of these motions occurs because of the variation of the vertical
gravitational acceleration (or, equivalently, the vertical frequency
$\Omega_z$, or the scaleheight $H$) with radius.  Vertical hydrostatic
equilibrium cannot be maintained in an eccentric disc because a fluid
element in an elliptical orbit experiences a vertical gravitational
acceleration that oscillates with the orbital frequency.  The
breathing mode is excited and the associated enthalpy perturbation
affects the horizontal dynamics, contributing to the precession of the
eccentric mode.  This contribution is found to be always
positive.  In the more general situation of a non-isothermal disc
undergoing adiabatic perturbations, the three-dimensional contribution
to $f$ (in which we compare a 3D-type theory with a 2DS-type theory)
is
\begin{equation}
  \f{3(\gamma+1)}{2\gamma}\f{P}{\Sigma r^2\Omega},
\end{equation}
where $\gamma$ is the adiabatic index and $P$ is the
vertically integrated pressure (Ogilvie \& Goodchild, in preparation),
but the derivation of this expression is beyond the scope of the
present paper.

\section{Full treatment}
\label{s:full}

\subsection{Inviscid dynamics}

The dynamical equations governing an isothermal, inviscid disc are
\begin{equation}
  (\p_t+\bu\bcdot\bnabla)\bu=-\bnabla(\Phi+h),
\label{motion}
\end{equation}
\begin{equation}
  (\p_t+\bu\bcdot\bnabla)h=-c_\rms^2\bnabla\bcdot\bu.
\end{equation}
When these are linearized about the basic state described in
Section~\ref{s:basic}, we obtain
\begin{equation}
  -\rmi\hat\omega u_r'-2\Omega u_\phi'=-\p_r h',
\end{equation}
\begin{equation}
  -\rmi\hat\omega u_\phi'+\f{\kappa^2}{2\Omega}u_r'=-\f{\rmi mh'}{r},
\end{equation}
\begin{equation}
  -\rmi\hat\omega u_z'=-\p_z h',
\end{equation}
\begin{eqnarray}
  \lefteqn{-\rmi\hat\omega h'+u_r'\p_r h+u_z'\p_z h}&\nonumber\\
  &&=-c_\rms^2\left[\f{1}{r}\p_r(ru_r')+\f{\rmi mu_\phi'}{r}+\p_zu_z'\right],
\end{eqnarray}
where $\hat\omega=\omega-m\Omega$ is the Doppler-shifted wave
frequency and the perturbations, denoted by primed quantities, have
the dependence $\exp(-\rmi\omega t+\rmi m\phi)$.

Following \citet{1987PASJ...39..457O}, we decompose the vertical
structure of the mode into the basis of Hermite polynomials defined by
\begin{equation}
  \he_n(\zeta)=\rme^{\zeta^2/2}\left(-\f{\rmd}{\rmd \zeta}\right)^n\rme^{-\zeta^2/2},
\end{equation}
where $\zeta=z/H$ is a dimensionless vertical coordinate and
$n=0,1,2,\dots$\  These polynomials satisfy the differential equation
\begin{equation}
  \he_n''(\zeta)-\zeta\,\he_n'(\zeta)+n\,\he_n(\zeta)=0,
\end{equation}
the recurrence relations
\begin{equation}
  \he_n'(\zeta)=n\,\he_{n-1}(\zeta),
\end{equation}
\begin{equation}
  \zeta\,\he_n(\zeta)=\he_{n+1}(\zeta)+n\,\he_{n-1}(\zeta)
\end{equation}
and the orthogonality relation
\begin{equation}
  \int_{-\infty}^\infty\rme^{-\zeta^2/2}\,\he_m(\zeta)\,\he_n(\zeta)\,\rmd \zeta=(2\pi)^{1/2}n!\,\delta_{mn}.
\end{equation}
The first three are $\he_0(\zeta)=1$, $\he_1(\zeta)=\zeta$ and
$\he_2(\zeta)=\zeta^2-1$.

We therefore expand
\begin{eqnarray}
  u_r'(r,z)&=&\sum_nu_n(r)\,\he_n(\zeta),\\
  u_\phi'(r,z)&=&\sum_nv_n(r)\,\he_n(\zeta),\\
  u_z'(r,z)&=&\sum_nw_n(r)\,\he_{n-1}(\zeta),\\
  h'(r,z)&=&\sum_nh_n(r)\,\he_n(\zeta).
\end{eqnarray}
with $u_n=v_n=h_n=0$ for $n<0$ and $w_n=0$ for $n<1$.  Bearing in mind
that $H$ depends on $r$, we have
\begin{equation}
  \p_r\,\he_n(\zeta)=-(\p_r\ln H)[n\,\he_n(\zeta)+n(n-1)\,\he_{n-2}(\zeta)].
\end{equation}
The projected equations are then
\begin{eqnarray}
  \lefteqn{-\rmi\hat\omega u_n-2\Omega v_n=-\p_r h_n}&\nonumber\\
  &&+(\p_r\ln H)[nh_n+(n+1)(n+2)h_{n+2}],
\label{un}
\end{eqnarray}
\begin{equation}
  -\rmi\hat\omega v_n+\f{\kappa^2}{2\Omega}u_n=-\f{\rmi mh_n}{r},
\label{vn}
\end{equation}
\begin{equation}
  -\rmi\hat\omega w_n=-\f{nh_n}{H},
\label{wn}
\end{equation}
\begin{eqnarray}
  \lefteqn{-\rmi\hat\omega\f{h_n}{c_\rms^2}+\f{1}{r\Sigma}\p_r(r\Sigma u_n)+\f{\rmi mv_n}{r}-\f{w_n}{H}}&\nonumber\\
  &&+(\p_r\ln H)(nu_n+u_{n-2})=0,
\label{hn}
\end{eqnarray}
\citep[cf.][]{2002ApJ...565.1257T,2006MNRAS.368..917Z}.

This approach corresponds to a (Galerkin) spectral treatment of the
partial differential equations governing the linearized dynamics,
which is much preferable to a finite-difference treatment.  In
practice this system of equations must be truncated by setting $u_n$,
etc., to zero for $n>N$ for some integer $N$.  The 2DO~theory is
obtained, in fact, by considering a radical truncation, $N=0$, of the
equations.

\subsection{Selected viscous effects}
\label{s:viscous}

To include viscosity, a term
\begin{equation}
  \f{1}{\rho}\bnabla\bcdot\bfT
\end{equation}
should be added to the right-hand side of the equation of
motion~(\ref{motion}), where
\begin{equation}
  \bfT=\rho\nu[\bnabla\bu+(\bnabla\bu)^\rmT]+\rho(\nu_\rmb-{\textstyle\f{2}{3}}\nu)(\bnabla\bcdot\bu)\bfone
\end{equation}
is the viscous stress tensor.  In the context of an isothermal disc it
is reasonable to assume that the kinematic shear and bulk viscosities
$\nu$ and $\nu_\rmb$ depend only on $r$.  We parametrize them as
\begin{equation}
  \nu=\alpha c_\rms H,\qquad\nu_\rmb=\alpha_\rmb c_\rms H.
\end{equation}

A full treatment of the effects of viscosity is complicated, not only
because the above expression for the viscous force must be evaluated
in cylindrical polar coordinates and then projected on to the basis of
Hermite polynomials, but also because the basic state is modified to
include a meridional flow driven by viscous stresses, which should be
considered in the linearized equations.  This problem is therefore
deferred to a future investigation.

In the present paper we adopt a simpler approach in which only
selected viscous effects are included.  We consider what might
be assumed to be the dominant viscous terms, i.e.\ those involving two
derivatives with respect to~$z$.  Since
\begin{equation}
  \f{1}{\rho}\p_z[\rho\p_z\,\he_n(\zeta)]=-\f{n}{H^2}\,\he_n(\zeta),
\end{equation}
the inviscid perturbation equations (\ref{un})--(\ref{wn}) are
modified by the addition of the viscous terms
\begin{equation}
  -\rmi\hat\omega u_n=\cdots-\f{\nu}{H^2}nu_n,
\label{unviscous}
\end{equation}
\begin{equation}
  -\rmi\hat\omega v_n=\cdots-\f{\nu}{H^2}nv_n,
\end{equation}
\begin{equation}
  -\rmi\hat\omega w_n=\cdots-\f{(\nu_\rmb+{\textstyle\f{4}{3}}\nu)}{H^2}(n-1)w_n,
\label{wnviscous}
\end{equation}
while equation~(\ref{hn}) is unchanged.  These terms act to damp the
mode, but have most effect on components of large $n$.  They have no
effect on $u_0$ and $v_0$, which represent horizontal motions
independent of $z$.  These viscous terms can also be thought of as
providing a coupling between different layers of the disc and thereby
encouraging it to adopt a horizontal motion independent of $z$.  We
show below that this effect is of considerable importance.

\subsection{Numerical solutions}

We solve the system of ordinary differential equations in $r$ for
modes with $m=1$ using a Chebyshev collocation (i.e.\ pseudospectral)
method.  This approach converts the differential equations and
boundary conditions into an algebraic generalized eigenvalue problem
for the frequency $\omega$, which we solve using a standard direct
method.  Specifically, equations~(\ref{un})--(\ref{hn}),
supplemented by the viscous terms
(\ref{unviscous})--(\ref{wnviscous}), are solved for $n=0,2,4,\dots,N$
with $u_n$, etc., set to zero for $n>N$.  Rigid boundary conditions
$u_n=0$ are adopted at both inner and outer boundaries, but it is
ensured that the modes obtained are completely insensitive to the
value of the outer radius and therefore to the choice of outer
boundary condition.

For comparison with the results in Section~\ref{s:powerlaw}, we
consider a disc with a midplane density profile $\rho_\rmm\propto
r^{-\sigma-3/2}$.  We also include a shear viscosity corresponding to
a constant $\alpha$ parameter, but no bulk viscosity.

Sample results are shown in Table~\ref{t:ot}.  The convergence of the
eigenfrequency with increasing truncation order $N$ of the Hermite
polynomial basis is remarkable.  The case $N=0$ corresponds exactly to
the two-dimensional theory considered by \citet{1991PASJ...43...75O},
and therefore agrees well with the 2DO~secular approximation.  Here
$\tilde Q=20$ is large enough to support a confined prograde mode.
The precession rate is much larger in the case $N=2$ and hardly varies
as further Hermite polynomials are included.  It agrees reasonably
well with the 3D~secular theory for an inviscid disc.  The slight
offset of the precession frequency is attributable partly to errors in
the secular approximation, which is valid only to leading order in
$\epsilon$, and partly to the effects of viscosity.  As described in
Appendix~\ref{s:appendix}, the viscous damping of vertical motions
considered in the full model can be represented within the 3D~secular
theory by multiplying the coefficient $9/4$ in the three-dimensional
expression~(\ref{f3d}) for $f$ by $(1-\rmi\beta)/(1+\rmi\beta)$, where
$\beta=\alpha_\rmb+{\textstyle\f{4}{3}}\alpha$.  Table~\ref{t:ot}
shows that this viscous secular theory gives good agreement with the
full model for $\alpha=0.1$.

\begin{table}
\caption{Scaled frequency eigenvalues obtained from the full
linearized equations for a disc with $\epsilon=0.02$, $\alpha=0.1$,
$\tilde Q=20$ and $\sigma=2$.  Rapid convergence is seen with
increasing values of the vertical truncation number $N$.  Comparable
results from the secular theories are shown below.  A negative value
of $\mathrm{Im}(\tilde\omega)$ represents the (scaled) exponential
damping rate of the mode.}
\begin{tabular}{lll}
Version&$\mathrm{Re}(\tilde\omega)$&$\mathrm{Im}(\tilde\omega)$\\
\hline
Full, $N=0$&1.189356&\\
Full, $N=2$&2.782638&-0.449908\\
Full, $N=4$&2.782585&-0.449510\\
Full, $N=6$&2.782585&-0.449510\\
\\
2DS secular, inviscid&0.636435&\\
2DO secular, inviscid&1.193322&\\
3D secular, inviscid&2.890299&\\
3D secular, viscous&2.829523&-0.448547\\
\hline
\end{tabular}
\label{t:ot}
\end{table}

The viscous damping rate of the modes is considerable.  Although the
dominant motion is horizontal and independent of $z$, and therefore
does not incur any viscous forces in our approximation, the
accompanying vertical motion is damped.  To excite eccentric modes in
a three-dimensional disc, this damping must be overcome.  Viscous
overstability may be able to do this, but detailed calculations are
required and there is uncertainty in the applicability of a
Navier--Stokes viscosity to turbulent stresses in the disc.  A
simple estimate can be made as follows.  In a two-dimensional,
isothermal, Keplerian shearing sheet with constant kinematic shear
viscosity $\nu$ and no bulk viscosity, the maximum local growth rate
of the overstability is $\approx0.034\,\alpha\Omega$ and occurs for a
radial wavelength $\approx13\,H$ \citep{2006MNRAS.372.1829L}.
Although our eigenfunctions do not have an obviously wavelike form,
this suggests that overstability may be able to compensate for the
damping rate found in Table~\ref{t:ot}, which corresponds to only
$\approx0.0018\,\alpha\Omega$ for the parameters adopted there.

A sufficiently large viscosity is required to couple different layers
in the disc effectively.  If the viscosity is reduced to
$\alpha=0.01$, with other parameters as in Table~\ref{t:ot}, a
frequency eigenvalue of $\tilde\omega=2.7831-0.1793\,\rmi$ is
obtained, and a slightly larger value of $N$ is required to obtain the
same convergence.  While the precession rate agrees well with the case
of $\alpha=0.1$, the damping rate is now larger than predicted by the
viscous secular 3D~theory.  This happens because of the increasing
$z$-dependence of the horizontal motion in the absence of a strong
coupling between layers; although $u_2$ is still much smaller than
$u_0$ for $\alpha=0.01$, the ratio $u_2/u_0$ is several times larger
than in the case $\alpha=0.1$.  The viscous damping of this shearing
motion is more important, for $\alpha=0.01$, than that of the vertical
motion considered in the viscous secular theory.

If $\alpha$ is reduced further, the precession rate starts to
deviate from the 3D~secular theory and more vertical structure
develops in the eigenfunction, requiring a larger value of $N$ for
convergence.  Similar behaviour was found by
\citet{2006MNRAS.372.1829L}.  The behaviour of a three-dimensional
inviscid eccentric disc could be very difficult to describe.

We also note that the stresses associated with tangled
magnetic fields in a disc in which the magnetorotational instability
occurs may provide an elastic, or viscoelastic, coupling between
different layers \citep{2001MNRAS.325..231O}.  The associated damping
rate may be smaller than that estimated on the basis of a
Navier--Stokes viscosity.

\section{Conclusions}
\label{s:conc}

In this paper we have examined the linear dynamics of one-armed
oscillation modes in the circumstellar discs of Be~stars.  A
three-dimensional effect, first identified by
\citet{2001MNRAS.325..231O} but neglected in previous treatments of
Be~stars, makes a crucial positive contribution to the precession
rates of such modes.  It allows confined prograde modes to be obtained
for all reasonable disc temperatures and stellar rotation rates.  This
property allows us to give a unified description of eccentric discs of
Be~stars of all stellar types without introducing uncertain radiative
forces or modifying the inner boundary condition.  While we do not
deny that in some cases radiative forces might be important, or that
the inner boundary condition might differ from a rigid one, we have
shown that these innovations are unnecessary.  We obtained these
results using a secular theory of eccentric discs and confirmed them
using a spectral treatment of the full linearized equations for
three-dimensional isothermal discs including viscous terms that couple
the horizontal motions at different altitudes.  In order to
make these modes grow, viscous damping must be overcome by an
excitation mechanism such as viscous overstability, which will be
investigated in a subsequent paper.

The three-dimensional dynamics that we have described may also
have important consequences for the behaviour of eccentric discs in
other circumstances.  For example, in cataclysmic variable stars
exhibiting superhumps, the relation between the precession rate of the
disc and the binary mass ratio is an important observational property
that is not fully explained by current theoretical models
\citep{2006MNRAS.368.1123G,2007MNRAS.378..785S}.

The physical model adopted in this paper is idealized.  To explain in
detail the observed cyclical behaviour of the emission lines of
Be~stars within this theoretical framework is likely to require a
treatment of non-isothermal and non-linear effects as well as a better
understanding of the time-dependent behaviour of the density
distribution of the circumstellar disc.  In addition, future work
should attempt to identify and assess mechanisms, such as viscous
overstability, by which one-armed oscillation modes may be excited in
circumstellar discs.  Nevertheless, the effect investigated in this
paper enhances the credibility of the one-armed oscillation model by
providing a natural explanation of confined prograde modes.

\section*{acknowledgments}

I thank Atsuo Okazaki, John Papaloizou and an anonymous
referee for useful comments.

\appendix

\section{Derivation of the secular theory}
\label{s:appendix}

We consider equations~(\ref{un})--(\ref{hn}) together with the viscous
effects described in Section~\ref{s:viscous}.  When $v_n$ and $w_n$
are eliminated, we have
\begin{eqnarray}
  \lefteqn{\f{(\hat\omega_{\mathrm{h}n}^2-\kappa^2)}{\rmi\hat\omega_{\mathrm{h}n}}u_n=-\p_rh_n+\f{2m\Omega}{r\hat\omega_{\mathrm{h}n}}h_n}&\nonumber\\
  &&+(\p_r\ln H)[nh_n+(n+1)(n+2)h_{n+2}],
\label{unelim}
\end{eqnarray}
\begin{eqnarray}
  \lefteqn{\left(-\rmi\hat\omega-\f{n\Omega_z^2}{\rmi\hat\omega_{\mathrm{v}n}}-\f{m^2c_\rms^2}{r^2}\f{1}{\rmi\hat\omega_{\mathrm{h}n}}\right)\f{h_n}{c_\rms^2}+\f{m\kappa^2}{2r\Omega\hat\omega_{\mathrm{h}n}}u_n}&\nonumber\\
  &&+\f{1}{r\Sigma}\p_r(r\Sigma u_n)+(\p_r\ln H)(nu_n+u_{n-2})=0,
\label{hnelim}
\end{eqnarray}
where $\hat\omega_{\mathrm{h}n}=\hat\omega+\rmi\alpha n\Omega_z$ and
$\hat\omega_{\mathrm{v}n}=\hat\omega+\rmi(\alpha_\rmb+{\textstyle\f{4}{3}}\alpha)(n-1)\Omega_z$
are the Doppler-shifted wave frequency modified to allow for the
viscous damping of horizontal and vertical motions, respectively.  We
are interested in the case $m=1$.

This system of equations is not closed because the $u_0$ equation
refers to $h_2$, which depends on $u_2$.  In turn, the $u_2$ equation
refers to $h_4$, which depends on $u_4$, and so on.  However, suppose
for the time being that $u_2$ can be neglected in the $h_2$ equation
(i.e.\ equation~\ref{hnelim} with $n=2$).  Then we have a
closed system
\begin{equation}
  \f{(\hat\omega^2-\kappa^2)}{\rmi\hat\omega}u_0=-\p_rh_0+\f{2\Omega}{r\hat\omega}h_0+2(\p_r\ln H)h_2,
\label{closed1}
\end{equation}
\begin{equation}
  \left(-\rmi\hat\omega-\f{c_\rms^2}{r^2}\f{1}{\rmi\hat\omega}\right)\f{h_0}{c_\rms^2}+\f{\kappa^2}{2r\Omega\hat\omega}u_0+\f{1}{r\Sigma}\p_r(r\Sigma u_0)=0,
\label{closed2}
\end{equation}
\begin{equation}
  \left(-\rmi\hat\omega-\f{2\Omega_z^2}{\rmi\hat\omega_{\mathrm{v}2}}-\f{c_\rms^2}{r^2}\f{1}{\rmi\hat\omega_{\mathrm{h}2}}\right)\f{h_2}{c_\rms^2}+(\p_r\ln H)u_0\approx0.
\label{closed3}
\end{equation}
These equations agree with a two-dimensional theory except for the
additional $h_2$ term in equation~(\ref{closed1}), which can be
related to $u_0$ through the approximate equation~(\ref{closed3}).

In the secular approximation we neglect the differences between
$\Omega$, $\kappa$, $\Omega_z$ and $\Omega_\rmK$ except where
essential, so that $\Omega\propto r^{-3/2}$ and $H\propto r^{3/2}$,
and we assume a low frequency $|\omega|\ll\Omega$.  The leading
approximations to the above equations in the inviscid case are then
\begin{equation}
  2\rmi\left(\f{\Omega^2-\kappa^2}{2\Omega}-\omega\right)u_0=-\p_rh_0-\f{2h_0}{r}+\f{3h_2}{r},
\end{equation}
\begin{equation}
  \rmi\Omega\f{h_0}{c_\rms^2}-\f{u_0}{2r}+\f{1}{r\Sigma}\p_r(r\Sigma u_0)=0,
\end{equation}
\begin{equation}
  -\rmi\Omega\f{h_2}{c_\rms^2}+\f{3u_0}{2r}=0.
\label{h2}
\end{equation}
These combine to give equation~(\ref{secular}) for $u=u_0$, with the
three-dimensional expression~(\ref{f3d}) for $f$.  If $h_2$ is
neglected altogether, equation~(\ref{secular}) is obtained, but with
the two-dimensional expression~(\ref{f2d}) for $f$.  It can be
seen from the above equations that $h_2$ always makes a positive
contribution to the precession rate $\omega$.

Neglecting $u_2$ compared to $u_0$ is equivalent to assuming that the
eccentricity is independent of $z$, since $\he_0(\zeta)=1$ and
$\he_2(\zeta)=\zeta^2-1$.  Under what conditions is this assumption
reasonable?  Rough estimates based on equations~(\ref{unelim})
and~(\ref{hnelim}) show that $u_2$ can be neglected in the $h_2$
equation if $\alpha\Omega$ is much larger than the precession
frequencies $\omega$ or $f$, meaning that the shear viscosity prevents
the significant development of a $z$-dependent horizontal motion.
This condition is readily satisfied in the circumstellar discs of
Be~stars for reasonable values of $\alpha$.  Ultimately, however, the
validation of this approximation comes from the numerical solution of
the full system of linearized equations.

If viscosity is retained in this analysis, within the secular
approximation, then the coefficient of $h_2$ in equation~(\ref{h2}) is
multiplied by $(1+\rmi\beta)/(1-\rmi\beta)$, where
$\beta=\alpha_\rmb+{\textstyle\f{4}{3}}\alpha$.  In this case,
equation~(\ref{secular}) is again obtained, but in
expression~(\ref{f3d}) for $f$ the coefficient $9/4$ of the
three-dimensional term is multiplied by $(1-\rmi\beta)/(1+\rmi\beta)$.
In this case the mode decays because of the viscous damping of the
vertical motion.  For sufficiently small $\alpha$, however,
the decay rate is enhanced by the viscous damping of the $z$-dependent
horizontal motion.

\label{lastpage}

\end{document}